# Magnetoelectric fields for microwave chirality discrimination in enantiomeric liquids

E. Hollander, E. O. Kamenetskii, and R. Shavit

Microwave Magnetic Laboratory,
Department of Electrical and Computer Engineering,
Ben Gurion University of the Negev, Beer Sheva, Israel

May 1, 2016

**Chirality discrimination is of a fundamental interest in biology, chemistry, and metamaterial studies. In optics, near-field plasmon-resonance spectroscopy with superchiral probing fields is effectively applicable for analyses of large biomolecules with chiral properties. We show possibility for microwave near-field chirality discrimination analysis based on magnon-resonance spectroscopy. Newly developed capabilities in microwave sensing using magnetoelectric (ME) probing fields originated from multiresonance magnetic-dipolar-mode (MDM) oscillations in quasi-2D yttrium-iron-garnet (YIG) disks, provide a potential for unprecedented measurements of chemical and biological objects. We report on microwave near-field chirality discrimination for aqueous *D*- and *L*-glucose solutions. The shown ME-field sensing is addressed to microwave biomedical diagnostics and pathogen detection and to deepening our understanding of microwave-biosystem interactions. It can be also important for an analysis and design of microwave chiral metamaterials.**

Many molecules in chemistry and biology are chiral. Biologically active molecules in amino acids (the building blocks of proteins) and sugars are chiral molecules. Chiral discrimination in a mixture of chiral molecules is among the most important and difficult tasks in biophysics and chemistry. In this connection, development of a technique that offers improved chiral analysis and better understanding of interactions of electromagnetic fields with chiral materials remains an important goal. Traditional chiroptical spectroscopy arises from the effect of interference between the electric-dipole transition moment and the weak magnetic-dipole transition moment that is detected when a chiral molecule is irradiated with alternating right- or left-circularly polarized light [1 – 3]. For localized (subwavelength) chiroptical biosensing, special plasmonic structures with left- and right-handed superchiral probing fields are effectively used [4 – 6].

The measured forms of chiroptical intensity are inversely proportional to the wavelength of the probing radiation. That is why use of microwave radiation to detect chirality was considered as a non-solvable problem. Surprisingly, a new microwave technique based on chirality-sensitive three-wave mixing to identify the enantiomers of chiral molecules, was demonstrated in Ref. [7]. This technique departs from traditional (optical) electromagnetic methods for detecting and identifying the handedness of molecules. Because it does not depend on a weak magnetic-dipole transition moment, the chiral signal is nearly as large as that of the applied microwaves. This conceptually new method to detect chirality is applicable, however, to cold gas-phase molecules. Based on the three-wave technique for molecules sampled in the gas phase [7], one cannot study handedness properties of liquid structures in microwaves, attractive for biological applications. Microwaves are attractive for biological applications because of their sensitivity to water and dielectric contrast. As a problem of great importance for biophysics and chemistry, this may concern, in particular, an analysis of the molecular mechanisms of nonthermal microwave effects [8, 9]. While microwave absorption is considered as an effective tool for observing and measuring different kinetic processes in biological structures, the existing standard microwave techniques are not applicable for sensing and monitoring enantiomeric liquids.

In this letter we report the first experimental observation of microwave chirality discrimination in liquid samples. In the room-temperature experiments, we use the aqueous *D*- and *L*- glucose solutions of different concentrations. Our microwave-spectroscopy technique determines the rotational energy levels of chiral molecules with the aim to be applied for localized measuring of different biological liquids and biological tissue. In literature, rotational spectroscopy for enantiomer-specific detection and separation is presented in many aspects. The Rabi frequency describing an electric dipole transition between rotational states of a chiral molecule differs in sign for opposite enantiomers [10, 11]. In the method used in Ref. [7], for identifying molecular chirality, the measured quantity depends on the handedness of three mutually orthogonal electric-dipole rotational transition moments, which are associated with the three rotational degrees of freedom of a molecule. When microwave radiation interacts with these moments, energy transfer changes the rotational state of the molecule, generating a spectroscopic signal. The sign of a scalar triple product depends on the order of the electric-dipole-moment vectors. This sign (being changed under spatial inversion and unchanged under time reversal) is a measure of molecule chirality. In 1978 Baranova and Zel'dovich [12] predicted the "propeller effect" in which a racemic mixture of chiral molecules is separated into left and right fractions when subjected to a radio frequency electric field of rotating polarization. The sense of rotation is given by the circularity of the electric field. Opposite enantiomers will 'screw' in opposite directions and thus separate along the direction about which the electric field rotates. In Ref. [13], it was shown that when exposed to a rotating electric field (at the frequency of 900 kHz), the left- and right-handed chiral molecules rotate with the field and act as microscopic propellers. Owing to their opposite handedness, they propel along the axis of field rotation in opposite directions. The results of chiral separation are detected by the chromatographic absorption profiles.

In our technique of microwave rotational spectroscopy, neither molecule rotational resonances nor "propeller effect" are used for enantiomer-specific detection. The main effect of chirality discrimination in our case is due to dependence of interaction between MDM resonances and microwave radiation on handedness properties of a sample loading a YIG disk. Similar to Ref. [7], a chiral particle is modeled as three mutually orthogonal electric-dipole moments and the handedness is identified by the sign of the triple product of these dipole moments. When a probe ME field originated from a MDM spectrum in a quasi-2D YIG disk is applied to a biological sample, electric dipoles of chiral molecules will have both spin and orbital angular momentums about a ferrite disk axis. The direction of molecule rotation is determined by the direction of a bias magnetic field. For a given direction of a bias field, transitions between rotational states of a chiral molecule are different for opposite enantiomers.

Recently, collective spins in YIG sub-millimeter spheres were explored to achieve their strong couplings to microwave radiation [14 – 16]. In such spheres, the couplings of the microwave photons to collective spin modes with non-uniform precession of magnetization – the MDMs [17, 18] – have also been observed. The spectrum of MDM oscillations in a quasi-2D YIG disk, however, is much stronger and richer than in a YIG sphere [19, 20]. Quantization of the microwave fields due to quasi-2D YIG disks with multi-resonance MDM oscillations appears as an interesting and unique effect [21, 22]. It was shown that a small quasi-2D ferrite disk with a MDM spectrum behaves as a source of specific quantized fields in vacuum termed magnetoelectric (ME) fields [23, 24]. The electric and magnetic fields of the ME fields carry both spin and orbital angular momentums. The ME fields originated from a MDM ferrite disk are strongly localized (subwavelength) fields with quantized topological characteristics, such as power-flow vortices and field helicities. Dependence of a sign of these topological parameters on a direction of a bias magnetic field allows distinguishing the "right" and "left" enantiomeric structures placed in close proximity to a MDM ferrite disk [24, 25].

The experimental set-up is sketched in Fig. 1. This is a microstrip structure with an embedded thin-film ferrite disk. For experimental studies, we use a ferrite sample with the following parameters. The YIG disk has a diameter of $D = 3$ mm and a thickness of $t = 0.05$ mm. The saturation magnetization of a ferrite is $4\pi M_s = 1880$ G. The linewidth of a ferrite is $\Delta H = 0.8$ Oe. The disk is normally magnetized by a bias magnetic field $H_0 = 4210$ Oe. The microstrip structure is realized on a dielectric substrate (Taconic RF-35 with the permittivity is $\varepsilon_r = 3.52$ and thickness of 1.52 mm). The characteristic impedance of a microstrip line is 50 Ohm. The S-matrix parameters were measured by a microwave network analyzer. With use of a current supply we established a quantity of a normal bias magnetic field $H_0$, necessary to get the MDM spectrum in a required frequency range. The reflection and transmission spectra of a microstrip structure with an unloaded MDM ferrite disk are the same as the spectra shown in Ref. [25]. At the MDM resonances, a ferrite disk becomes slightly entangled in the reflected microwave radiation, while becomes strongly entangled in the transmitted microwave radiation. For this reason, one observes Lorentz-type resonances in the reflection spectrum and Fano interference in the transmission response.

In the present experiment, a ferrite disk is loaded by a cylindrical capsule with 15 μl capacitance of aqueous D- or L-glucose solutions. To avoid strong damping of MDM resonances by a load, the capsule is placed with a gap of 300 um above a ferrite disk. Figure 2 shows experimental results of the MDM transmission spectra in a microstrip structure for the D- or L-glucose solutions at two opposite directions of a bias magnetic field. In both cases of the solutions, the glucose concentration is 538 mg/ml. For better illustration of the effect, the spectra are normalized to the background (when a bias magnetic field is zero) level of the microwave structure. The results give evidence for a specific chiral symmetry: simultaneous change of the glucose handedness and direction of bias of the magnetic field keeps the system symmetry unbroken.

For a quantitative analysis of chirality discrimination in glucose solutions, we use the MDM reflection spectra in a microstrip structure. The spectra are normalized to the background level of the microwave structure similar to the normalization used for the transmission spectra. The reflection spectra for the D- or L-solutions at two opposite directions of a bias magnetic field are shown in Fig. 3 for the concentration of 538 mg/ml for the two types of glucose solutions. Both in Fig. 2 and in Fig. 3 the resonance peaks are classified as the radial and azimuthal modes [21]. For the quantitative analysis, we put the first radial resonance peaks of all the spectra at the same frequency. For two types of the glucose solutions (right- and left-handed) and two opposite directions of the first peak, this frequency was 7.726 GHz. Based on the results in Fig. 3, we calculated for every type of a glucose solution the differences between the scattered-matrix parameters obtained at two opposite directions of a bias magnetic field:

$$\Delta S = S_{21}\left(H_0^{(+)}\right) - S_{21}\left(H_0^{(-)}\right), \tag{1}$$

where signs (±) indicate the bias magnetic field directions along a disk axis. The enantiomer-dependent results are shown in Fig. 4. It is evident that for all the resonant peaks the quantities of $\Delta S$ have definite antisymmetric forms with respect to frequency. These antisymmetric forms are oppositely oriented for the D- and L-glucose. A quantitative characteristic of the sample chirality is made by estimation of the frequency differences $\Delta f$ between the peaks of parameters $\Delta S$. We can see that for different types of MDMs such frequency differences are different. In the present study, we made a relative estimation of the liquid chirality via variation of concentration of the glucose solution. Fig. 5 shows the results of enantiomer-dependent parameters $\Delta f$ for two concentrations, 420 mg/ml and 538 mg/ml, of the glucose solutions.

To explain the observed experimental results of microwave chirality discrimination in enantiomeric liquids we analyze a theoretical model. This model, based on ME properties of MDM oscillations and on interaction of the ME fields with matter, arises from the symmetry analysis. Symmetry principles play an important role with respect to the laws of nature. Maxwell added an electric displacement current to put into a symmetrical shape the equations coupling together the electric and magnetic fields. The dual symmetry between electric and magnetic fields underlies the conservation of energy and momentum for electromagnetic fields. This symmetry underlies also the conservation of optical (electromagnetic) helicity [26, 27]. In a small ferrimagnetic sample [with sizes much less than the free-space electromagnetic (EM) wavelength], one has negligibly small variation of electric energy and microwave dynamical processes are described by "incomplete" Maxwell equations with neglect of an electrical displacement current [18, 28]. However, in spite of breaking of Maxwell's symmetry, specific unified-field (with coupled electric- and magnetic-field components) retardation effects exist in such a sample. There are MDM oscillations. At the MDM resonances, both the electric and magnetic fields inside and in close proximity outside a ferrite disk have spin and orbital angular momentums. These rotating fields are not mutually perpendicular. The fields originated from MDM oscillations – the ME fields – are different from electromagnetic fields.

The quadratic forms characterizing ME fields are power-flow and helicity densities. The ME-field helicity density is defined as [24] $F = \frac{\varepsilon_0}{2} \vec{E} \cdot \nabla \times \vec{E}$. Formally, this parameter can be considered as a particular case of the EM-field chirality [26] $\chi = \frac{\varepsilon_0}{2} \vec{E} \cdot \nabla \times \vec{E} + \frac{1}{2\mu_0} \vec{B} \cdot \nabla \times \vec{B}$, when an electrical displacement current is negligibly small. At the MDM resonances ($\omega = \omega_{MDM}$), one observes active and reactive power flows [29]. The active power flow has vortex topology. The helicity parameter is calculated as

$$F = \frac{\omega_{MDM} \varepsilon_0}{4} \text{Re}(\vec{E} \cdot \vec{B}^*) = \frac{\omega_{MDM}}{4c^2} \text{Re}(\vec{E} \cdot \vec{H}^*), \qquad (2)$$

where $c$ is the light velocity in vacuum. A sign of the helicity parameter depends on a direction of a bias magnetic field: $F^{\vec{H}_0 \uparrow} = -F^{\vec{H}_0 \downarrow}$. An integral of the ME-field helicity over the entire near-field vacuum region of volume $V$ should be equal to zero:

$$\mathcal{H} = \int_V F dV = \frac{\omega_{MDM}}{4c^2} \int_V \text{Re}(\vec{E} \cdot \vec{H}^*) dV = 0. \qquad (3)$$

The helicity parameter $F$ is a pseudoscalar: to come back at the initial stage, one has to combine a reflection in a ferrite-disk plane and an opposite (time-reversal) rotation about an axis perpendicular to that plane. This property is illustrated in Fig. 6.

The MDM spectral solutions [23, 24] show topologically robust field structures which rotate around a disk axis. We have both sinusoidal electric and magnetic fields propagating in the azimuthal direction. At the MDM resonance, the frequency of the orbital rotation of the fields is the same as the frequency of the magnetization precession in YIG. In a laboratory frame, one observes not only magnetization, but also electric-polarization properties of a ferrite disk. The spin-orbit interaction, resulting in electric polarization for magnetic-dipole dynamics, produces a rotating electric field which is not perpendicular to a rotating magnetic field. Since the electric dipoles rotate at the same frequency as the magnetic dipoles, an electric displacement current is still negligibly small. At the MDM resonances, the rotating electric fields cause rotation of electric dipoles of chiral molecules. It was shown in Ref. [30] that there can be not one, but an infinite number of chiral parameters that characterize a chiral object and each of these chiral parameters is a pseudoscalar. The vector triple product of electric dipoles, modeling a chiral molecule $\vec{d}_1 \cdot (\vec{d}_2 \times \vec{d}_3)$, is a simple case of such a pseudoscalar. This pseudocalar has opposite signs for the right and left chiral molecules. The helicity parameter of a ME field is a pseudoscalar as well. So the field interacting with chiral molecules is distinguished by the same topological property (see Fig. 6). When a dielectric sample is placed above a ferrite disk, every electric dipole in a sample will rotate. The direction of rotation is determined by the direction of a bias magnetic field. The field chirality results in unidirectional transfer of angular momenta through a subwavelength vacuum region. For a given direction of a bias field, transitions between rotational states of a chiral molecule are different for opposite enatiomers. One can effectively distinguish "right" and "left" enantiomeric properties of samples.

In this letter we demonstrated a conceptually new form of spectroscopy of enantiomeric liquids based on a microwave technique. The technique determines chiral properties of molecules with the aim to be applied for localized measuring of different biological liquids and biological tissue at room temperature.

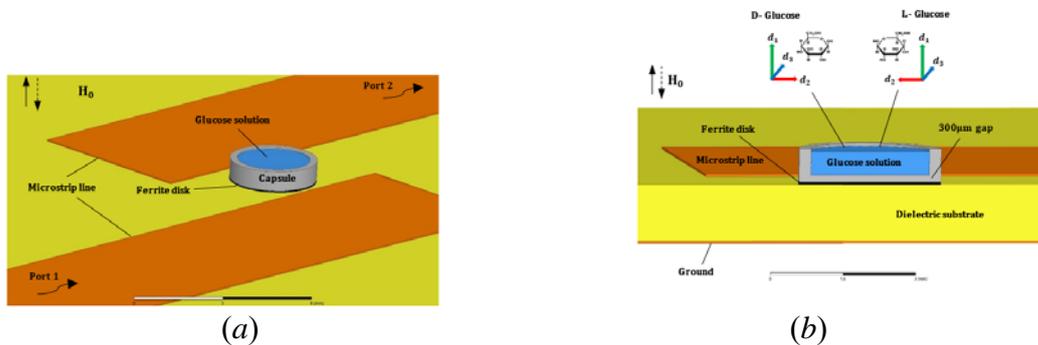

(*a*)                 (*b*)

Fig. 1. A microstrip structure with an embedded thin-film ferrite disk. A ferrite disk is loaded by a cylindrical capsule with aqueous *D*- or *L*-glucose solutions. To avoid strong damping of MDM resonances by a load, the capsule is placed with a gap of 300 um above a ferrite disk.

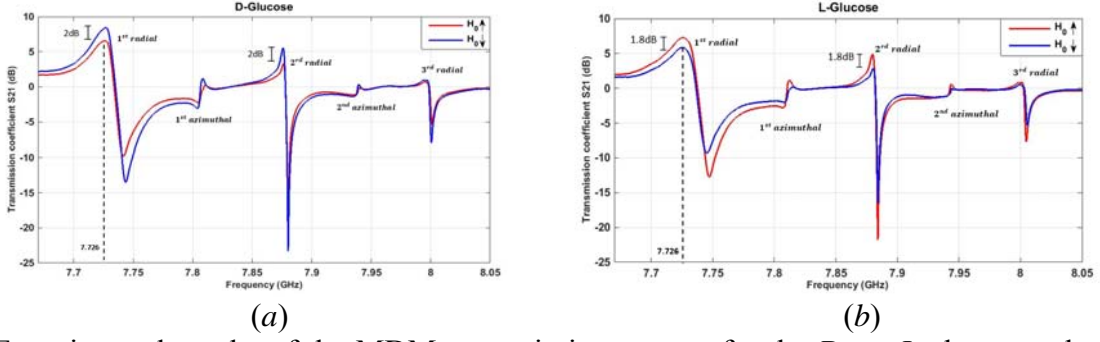

(*a*)          (*b*)

Fig. 2. Experimental results of the MDM transmission spectra for the *D*- or *L*-glucose solutions at two opposite directions of a bias magnetic field. In both cases of the solutions, the glucose concentration is 538 mg/ml. The spectra are normalized to the background (when a bias magnetic field is zero) level of the microwave structure. The results give evidence for a specific chiral symmetry: simultaneous change of the glucose handedness and direction of bias of the magnetic field keeps the system symmetry unbroken. The resonance peaks are classified as the radial and azimuthal modes [21]. The frequency $f$ = 7.726 GHz is the resonance frequency of the 1$^{st}$ radial MDM.

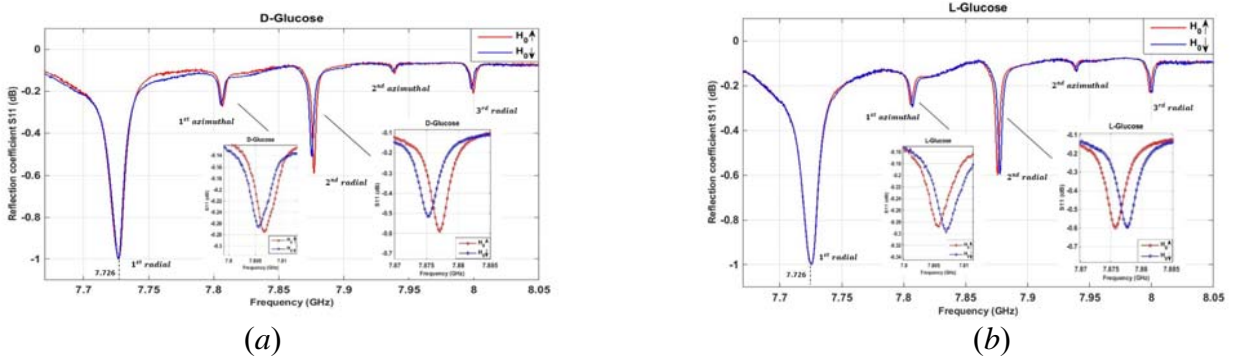

(*a*)          (*b*)

Fig. 3. The MDM reflection spectra for the *D*- or *L*-solutions at two opposite directions of a bias magnetic field for the concentration of 538 mg/ml for the two types of glucose solutions. The spectra are normalized to the background level of the microwave structure similar to the normalization used for the transmission spectra. The resonance peaks are classified as the radial and azimuthal modes [21]. The frequency $f$ = 7.726 GHz is the resonance frequency of the 1$^{st}$ radial MDM.

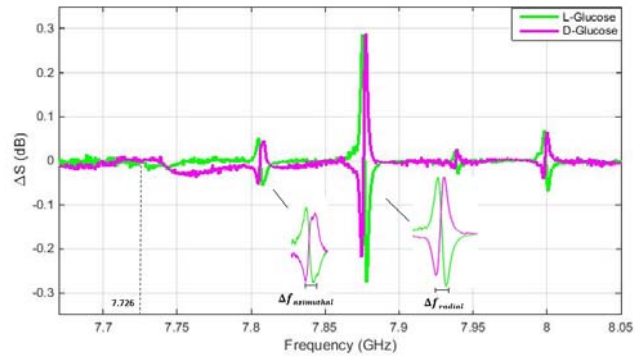

Fig. 4. The enantiomer-dependent parameter $\Delta S = S_{11}\left(H_0^{(+)}\right) - S_{11}\left(H_0^{(-)}\right)$ obtained at two opposite directions of a bias magnetic field for the *D*- and *L*-glucose solutions. For all the MDM resonant peaks the quantities of $\Delta S$ have definite antisymmetric forms with respect to frequency. These antisymmetric forms are oppositely oriented for the *D*- and *L*-glucose. A quantitative

characteristic of the sample chirality is made by estimation of the frequency differences $\Delta f$ between the peaks of parameters $\Delta S$ for the *D*- and *L*-glucose solutions. For different MDMs the frequency differences $\Delta f$ are different.

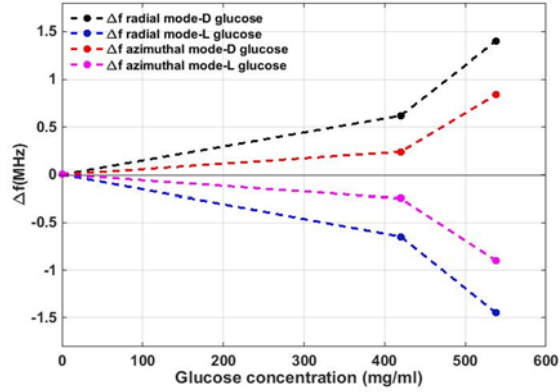

Fig. 5. Enantiomer-dependent parameters $\Delta f$ for two concentrations, 420 mg/ml and 538 mg/ml, of the *D*- and *L*- glucose solutions.

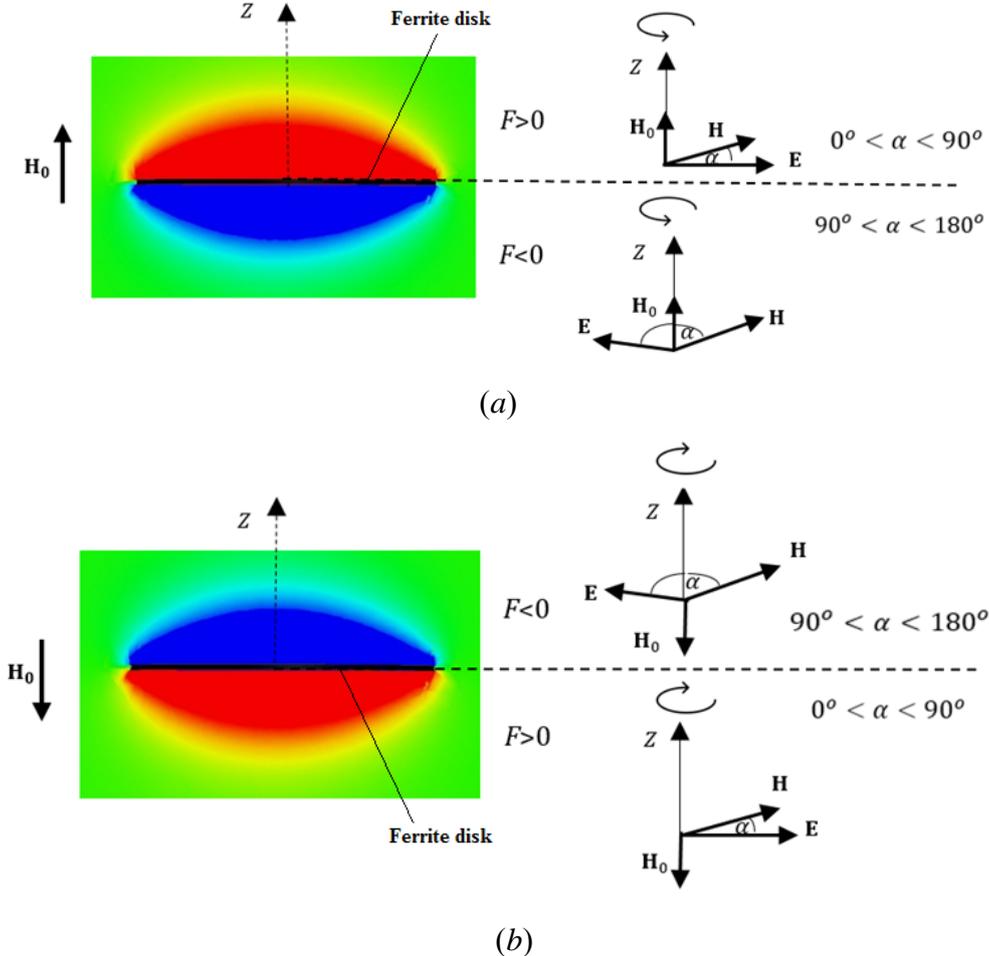

Fig. 6. The field topology near a ferrite disk at the MDM resonance frequency and at two opposite directions of a bias magnetic field. The electric and magnetic fields outside a ferrite disk are rotating fields which are not mutually perpendicular. The helicity parameter $F$ is a pseudoscalar: to come back at the initial stage, one has to combine a reflection in a ferrite-disk plane and an opposite (time-reversal) rotation about an axis perpendicular to that plane.

**METHODS AND SUPPLEMENTARY INFORMATION**

The magnetic insulator yttrium iron garnet $Y_3Fe_5O_{12}$ (YIG) provides an ideal platform for the study of spin waves. Long range dipole-dipole correlation in a ferrimagnetic sample can be treated in terms of collective excitations of the system as a whole. Ferrite samples with linear dimensions smaller than the dephasing length, but still much larger than the exchange-interaction scales, are mesoscopic structures with magneto-dipolar-mode (MDM) oscillations. Recently, it was shown that mesoscopic quasi-2D ferrite disks, distinguishing by multiresonance MDM spectra, demonstrate unique properties of artificial atomic structures: energy eigenstates, eigen power-flow vortices, and eigen helicity parameters. Because of these properties, MDMs in a ferrite disk enable the confinement of microwave radiation to subwavelength scales. In microwave structures with embedded MDM ferrite samples, one can observe quantized fields with topologically distinctive characteristics. These fields are termed magnetoelectric (ME) fields [1 – 7].

At the MDM resonance, both the electric and magnetic fields inside and in close proximity outside a ferrite disk are the fields sinusoidal propagating in the azimuthal direction. Fig. 1 shows the field structures of the 1st radial MDM at the upper surface of a ferrite disk. The fields have spin and orbital angular momentums. The spin-orbit interaction results in electric polarization for magnetic-dipole dynamics and magnetization for electric-dipole dynamics. When a ferrite disk is loaded by a capsule with glucose solution, interaction of the MDM rotating electric field with chiral molecules causes changes in magnetization dynamics in a ferrite disk, which, in its turn, causes changes in interaction of MDM oscillations with electromagnetic fields of a microwave structure. This results in transformation of the transmission/reflection characteristics of a microwave structure. The transmission/reflection coefficients were measured with use of a microwave network analyzer. With the use of a current supply we established a quantity of a normal bias magnetic field, necessary to get the MDM spectrum in a required frequency range. By switching the current direction, we changed a direction of a bias magnetic field. Since we put the first radial resonance peaks of all the spectra at the same frequency, the hysteresis effects in a magnetic system at switching the current direction were eliminated. In our experiments, we assume that electric field rotation is much slower than the rotational relaxation time of the molecules in low-viscosity solvents (typically in the 100-500 ps timescale [8, 9]).

To illustrate the effect of interaction of MDM oscillations with chiral structures, we use numerical simulations for microwave near-field chirality discrimination of a planar chiral structure. Recently, the effect of interaction of microwave radiation with a planar chiral structure patterned on a subwavelength scale was studied experimentally for two enantiomeric forms [10]. It was found that this is a polarization sensitive transmission effect asymmetric with respect to the direction of wave propagation. Our technique to recognize the handedness of a planar chiral structure is completely different. Based on our near-field technique, we study numerically chiral metamaterials realized as a composition of the "right" and "left" planar metallic stars. Fig. 2 shows a rectangular waveguide (cross-section sizes are 22.86 mm × 10.16 mm) with an embedded MDM ferrite disk. A two-dimensional metamaterial composed by the "right" and "left" planar metallic elements is placed above a ferrite disk. Fig. 3 shows numerical results of the MDM transmission spectra for the right- and left-handed chiral metamaterials at two opposite directions of a bias magnetic field. The results give evidence for a specific chiral symmetry: simultaneous change of the metamaterial handedness and direction of bias of the magnetic field keeps the system symmetry unbroken. Quantitatively, chirality discrimination in metamaterials is analyzed based on the MDM reflection spectra for the right- and left-handed metamaterials and at two opposite directions of a bias magnetic field. These spectra are shown in Fig. 4. For the quantitative analysis, we put the first radial resonance peaks of all the spectra at the same frequency. For two types of the metamaterials and two opposite directions of the first peak, this frequency was 8.522 GHz. The enantiomer-dependent parameter $\Delta S = S_{21}\left(H_0^{(+)}\right) - S_{21}\left(H_0^{(-)}\right)$ obtained at two opposite directions of a bias magnetic field for the right- and left-handed metamaterials is shown in Fig. 5. For all the MDM resonant peaks the quantities of $\Delta S$ have definite antisymmetric forms with respect to frequency. These antisymmetric forms are oppositely oriented for the right- and left-handed metamaterials. A quantitative characteristic of the sample chirality is made by estimation of the frequency differences $\Delta f$ between the peaks of parameters $\Delta S$. For different MDMs the frequency differences $\Delta f$ are different.

In the metamaterial structures, the azimuthally rotating electric and magnetic fields of a MDM ferrite disk induce electric charges and currents on metal stars. The direction of the field rotation is determined by the direction of a bias magnetic field. For a given enantiomeric structure, the electric charges and currents induced on metal stars are different for different directions of a bias magnetic field. Fig. 6 shows the field topologies in a metamaterial with given chirality at two opposite directions of a bias magnetic field. There are the power-flow density distributions and distributions of the helicity parameter. One can see that for given chirality of a metamaterial, the handed metal stars are faced with different power flows at opposite directions of a bias magnetic field. The opposite (time-reversal) rotations about a disk axis give evidence for the pseudoscalar properties of the helicity parameter in a plane of metamaterial structure.

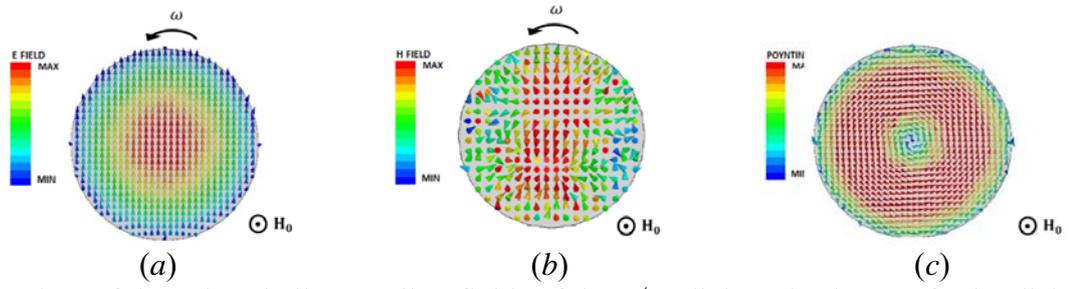

Fig. 1. A top view of the azimuthally traveling fields of the 1st radial mode above a ferrite disk, at a given direction of a bias magnetic field. (*a*) Electric field; (*b*) magnetic field. (*c*) The power-flow density vortex.

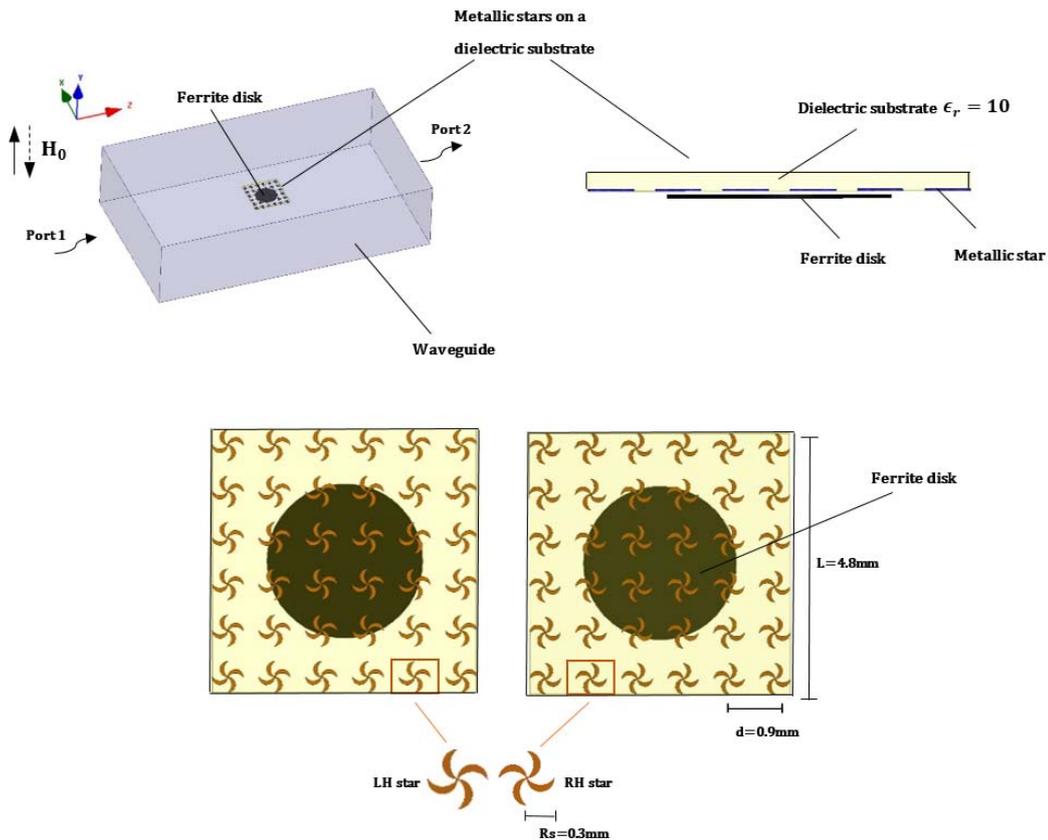

Fig. 2. A rectangular waveguide with an embedded MDM ferrite disk. A two-dimensional metamaterial composed by the "right" and "left" planar metallic elements is placed above a ferrite disk.

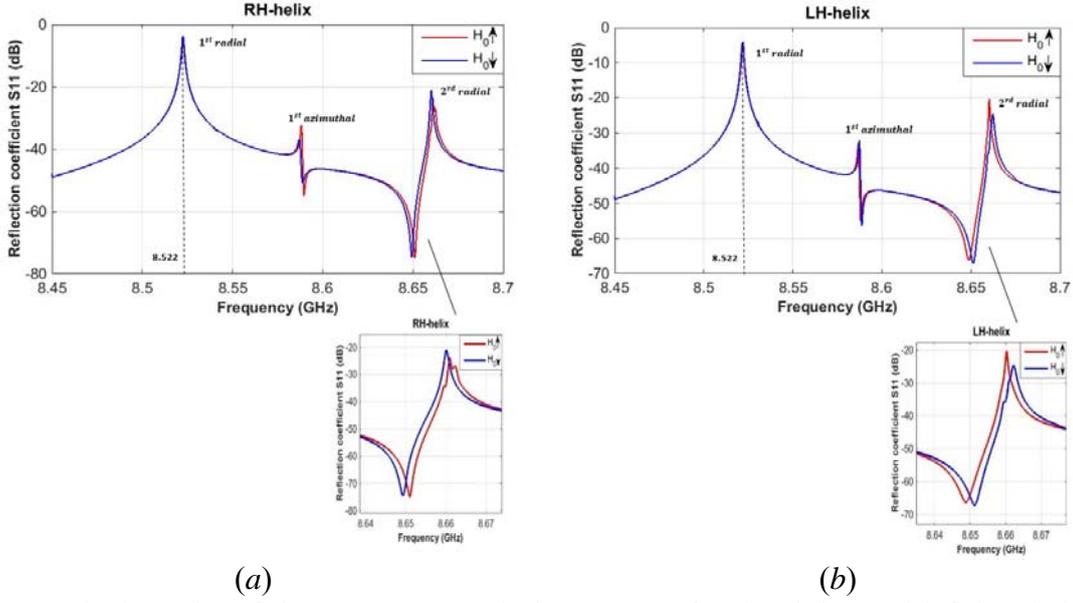

(*a*)           (*b*)

Fig. 3. Numerical results of the MDM transmission spectra for the right- and left-handed chiral metamaterials at two opposite directions of a bias magnetic field. The resonance peaks are classified as the radial and azimuthal modes. The frequency $f = 8.522$ GHz is the resonance frequency of the 1$^{st}$ radial MDM. The results give evidence for a specific chiral symmetry: simultaneous change of the metamaterial handedness and direction of bias of the magnetic field keeps the system symmetry unbroken.

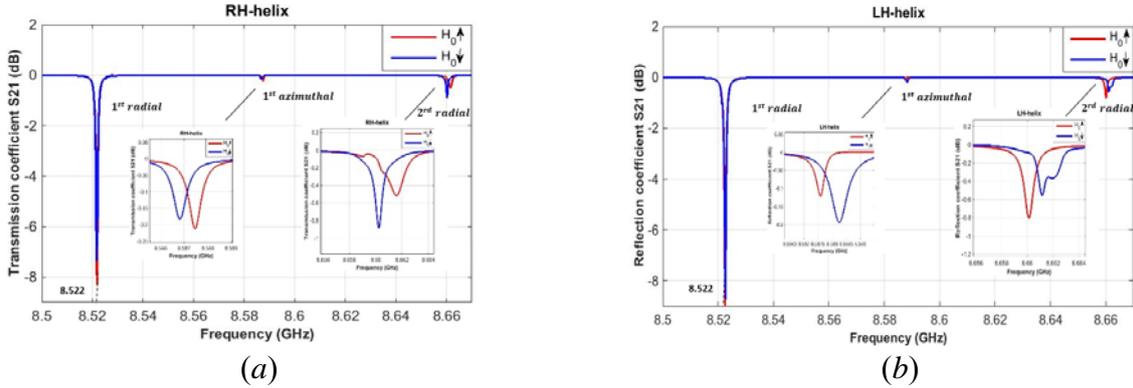

(*a*)           (*b*)

Fig. 4. The MDM reflection spectra for the right- and left-handed metamaterials at two opposite directions of a bias magnetic field. The resonance peaks are classified as the radial and azimuthal modes. The frequency $f = 8.522$ GHz is the resonance frequency of the 1$^{st}$ radial MDM.

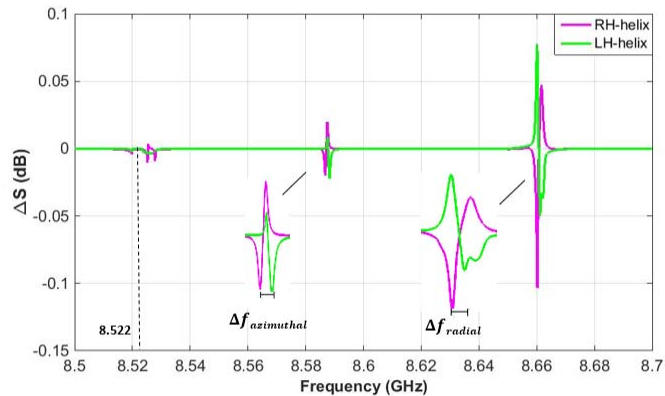

Fig. 5. The enantiomer-dependent parameter $\Delta S = S_{21}\left(H_0^{(+)}\right) - S_{21}\left(H_0^{(-)}\right)$ obtained at two opposite directions of a bias magnetic field for the right- and left-handed metamaterials. For all the MDM resonant peaks the quantities of $\Delta S$ have definite antisymmetric forms with respect to frequency. These antisymmetric forms are oppositely oriented for the right- and left-handed metamaterials. A quantitative characteristic of the sample chirality is made by estimation of the frequency differences $\Delta f$ between the peaks of parameters $\Delta S$. For different MDMs the frequency differences $\Delta f$ are different.

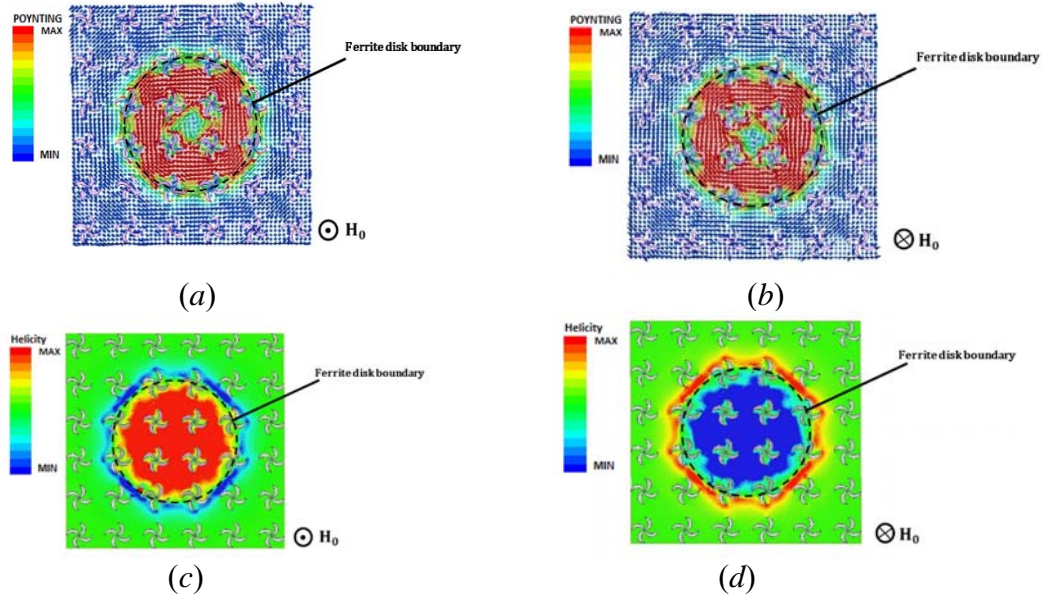

(*a*)  (*b*)

(*c*)  (*d*)

Fig. 6. The field topologies in a metamaterial with given chirality at two opposite directions of a bias magnetic field. (*a*), (*b*) the power-flow density distribution; (*c*), (*d*) distributions of the helicity parameter *F* in a plane of metamaterial structure.